\documentclass[twocolumn,preprintnumbers,10pt,floatfix,nofootinbib]{revtex4} 

\long\def\del #1 \enddel { }

\usepackage{graphicx}
\usepackage{amsmath}
\usepackage{amssymb}
\usepackage{subfigure}

\usepackage{epsfig}

\begin{document}

\def\beq{\begin{equation}}
\def\eeq{\end{equation}}

\def\bea{\arraycolsep .1em \begin{eqnarray}}
\def\eea{\end{eqnarray}}
\def\Tr{{\rm Tr}}

\def\dq{{q \llap{/}}\,}
\def\dk{{k \llap{/}}\,}
\def\step{\\[-2ex]}

\let\de=\delta
\let\Ga=\Gamma
\let\eps=\epsilon
\let\om=\omega
\def\bpi{\mbox{\boldmath$\pi$}}
\def\bphi{\mbox{\boldmath$\phi$}}

\def\Eq#1{Eq.~(\ref{#1})}
\def\eq#1{(\ref{#1})}
\def\fig#1{fig.~(\ref{#1})}

\def\s0#1#2{\mbox{\small{$ \frac{#1}{#2} $}}}
\def\0#1#2{\frac{#1}{#2}}

\def\llangle{\left\langle}
\def\rrangle{\right\rangle}

\title{High-accuracy scaling exponents in the local potential approximation}

\author{Claude Bervillier${}^{a}$} \author{Andreas J\"uttner${}^{b}$}
\author{Daniel F. Litim${}^{c,d}$} \address{\mbox{${}^a$ Laboratoire
de Math\'ematiques et Physique Th\'eorique, Universit\'e de Tours, Parc
de Grandmont, F-37200 Tours}\\ \mbox{${}^b$ School of Physics and
Astronomy, University of Southampton, Southampton SO17 1BJ, U.K.}\\
\mbox{${}^c$ Department of Physics and Astronomy, University of
Sussex, Brighton, East Sussex, BN1 9QH, U.K.}\\ \mbox{${}^d$ Theory
Group, Physics Division, CERN, CH-1211 Geneva 23.}}
\begin{abstract}
  We test equivalences between different realisations of Wilson's
  renormalisation group by computing the leading, subleading, and
  anti-symmetric corrections-to-scaling exponents, and the full fixed point
  potential for the Ising universality class to leading order in a derivative
  expansion.  We discuss our methods with a special emphasis on accuracy and
  reliability.  We establish numerical equivalence of Wilson-Polchinski flows
  and optimised renormalisation group flows with an unprecedented accuracy in
  the scaling exponents.  Our results are contrasted with high-accuracy
  findings from Dyson's hierarchical model, where a tiny but systematic
  difference in all scaling exponents is established.  Further applications
  for our numerical methods are briefly indicated.
\end{abstract}
\preprint{CERN-PH-TH-2007-006, SHEP-0702}

\maketitle

\section{Introduction}
Renormalisation group (RG) methods provide powerful tools in the computation
of universal scaling exponents or anomalous dimensions in quantum field
theories \cite{Zinn-Justin:1989mi}. Modern functional RG approaches are based
on Wilson's idea to integrate-out momentum modes within a path-integral
representation of the theory~\cite{Bagnuls:2000ae}. A particular strength of
these methods is their flexibility when it comes to approximations. Wilsonian
flows can be implemented in many different ways, which helps adapting the
technique to the problem at hand \cite{Pawlowski:2005xe}.
\step

A useful systematic expansion scheme is the derivative expansion, which, to
leading order, is known as the local potential approximation (LPA). For the
study of critical phenomena, the derivative expansion is expected to provide
good results as long as quantum corrections to propagators and anomalous
dimensions remain small~\cite{Golner:1986}.  Important results in LPA have
been accumulated over the years based on different implementations of the RG
including the Wilson-Polchinski RG \cite{Wilson:1973jj,Polchinski:1983gv}, the
functional RG \cite{Bagnuls:2000ae} for the effective average action
\cite{continuum} and optimised versions thereof \cite{Litim:2001up}, and
Dyson's hierarchical RG \cite{Dyson:1968up,Baker:1972}. Interestingly, and
despite qualitative differences in the respective formalisms, these approaches
are closely related in LPA. In fact, it has been established by Felder
\cite{Felder:1987} that the Wilson-Polchinski RG and the hierarchical RG are
equivalent. More recently, following a conjecture first stated in
\cite{Litim:2001fd,Litim:2005us}, the equivalence between the
Wilson-Polchinski flow and an optimised functional RG has been proven by
Morris \cite{Morris:2005ck}.  Consequently, universal scaling exponents from
either of the three approaches should be identical.\step

Quantitatively, these equivalences have been put to test for $3d$ critical
scalar theories \cite{Litim:2005us,Godina:2005hv}. While scaling exponents
from either approach agree at the order $10^{-4}$, an unexpected disagreement
starting at the order $10^{-5}$ was found. Presently, however, this last
observation solely relies on results from the hierarchical RG and the
optimised RG, where exponents have been obtained with sufficiently high
accuracy. Therefore it becomes mandatory to obtain results from the
Wilson-Polchinski RG with a higher precision, in order to confirm or refute
the above-mentioned discrepancy. \step

In this paper, we close this gap in the literature and study both the
Wilson-Polchinski RG and the optimised RG for the 3d Ising universality class
with an unprecedented accuracy. To that end, we introduce several methods to
solve non-linear eigenvalue problems with special emphasis on the numerical
reliability. We expect that these techniques are equally useful for other
non-linear problems in mathematical physics.  The outline of the paper is as
follows. We introduce the relevant differential equations (Sec.~\ref{RGs}) and
discuss our numerical methods (Sec.~\ref{Numerical}) and the error
control~(Sec.~\ref{Error}).  Results for the fixed point solution and scaling
exponents are discussed (Sec.~\ref{Results}) and compared with the
hierarchical model (Sec.~\ref{Comparison}). We close with a discussion and
conclusions (Sec.~\ref{Conclusions}).

\section{Local potential approximation}\label{RGs}
We restrict ourselves to a real scalar field theory in three dimensions -- the
Ising universality class -- and to the leading order in the derivative
expansion, meaning that the scale-dependent effective action is approximated
by $\Gamma_k=\int d^3x [\s012 \partial_\mu\phi\partial_\mu\phi +V_k(\phi)]$.
Here, $k$ denotes the scale of the Wilsonian momentum cutoff. We provide the
flows in terms of the dimensionless potentials $u(\rho)=V_k(\phi)/k^3$ with
$\rho=\s012\phi^2/k$, and $v(\varphi)=V_k(\phi)/k^3$ with
$\varphi=\phi/\sqrt{k}$.
Within an effective average action approach \cite{continuum}, the flow
equation for $u(\rho)$ depends on the infrared momentum cutoff. For an
optimised choice of the latter \cite{Litim:2001up}, and neglecting an
irrelevant (field-independent) term, it is given by
\begin{equation}
\label{flow}
\partial_t u=
-3u+ \rho u' +\frac{1}{1+u'+2\rho\,u''}\,.
\end{equation}
Here, $t=\ln k$ denotes the logarithmic scale parameter. At a fixed point
$\partial_tu'=0$, the potential obeys
\begin{equation}
\label{FP-opt}
\frac{{\rm d}u''}{{\rm d}\rho}=
\left(\frac{u''}{2}-\frac{u'}{\rho}\right)
(1+u'+2\rho\,u'')^2 - \frac{3u''}{2\rho}\,.
\end{equation}
There exists a unique non-trivial and well-defined (finite, no poles) solution
to \eq{FP-opt} which extends over all fields. In terms of the potential
$v(\varphi)$, the corresponding equations are
\begin{eqnarray}
\label{flow2}
\partial_t v&=&
-3v+ \012\varphi\, v' +\frac{1}{1+v''}\,,\\
\label{FP-opt2}
\frac{{\rm d}v''}{{\rm d}\varphi}&=&
\012\left(\varphi\,v''-5v'\right)(1+v'')^2\,.
\end{eqnarray}
For the Wilson-Polchinski RG, we follow an analogous ansatz for the
effective action. In terms of the potential $u(\rho)$ and with $t=\ln k$ we
find
\begin{eqnarray}
\label{flowPolchinski}
\partial_t u&=&
-3u+ \rho u'-u'-2 \rho\,u''+2\rho\,(u')^2\\
\label{FP-Polchinski}
\frac{{\rm d}u''}{{\rm d}\rho}&=&
\frac{u''}{2}-\frac{u'}{\rho}
+\frac{(u')^2}{\rho}+2u'\,u''-\frac{3u''}{2\rho} \,.
\end{eqnarray}
The corresponding equations for $v(\varphi)$ are
\begin{eqnarray}
\label{flowPolchinski2}
\partial_t v&=&
-3v+ \012\varphi\,v'-v''+(v')^2 \,,\\
\label{FP-Polchinski2}
\frac{{\rm d}v''}{{\rm d}\varphi}&=&
\012\left(\varphi\,v''-5v'\right) + 2v'\,v'' \,.
\end{eqnarray}
Numerical solutions for the fixed point potentials and their derivatives
\eq{FP-opt}, \eq{FP-opt2} and \eq{FP-Polchinski2} are displayed in
Figs.~\ref{fUs},~\ref{fUprime},~\ref{fUprime2},~\ref{pComparisonSmallField}
and~\ref{pComparisonLargeField} below.\step

The potential $v(\varphi)$ in \eq{flow2} is related to the Wilson-Polchinski
potential $v_{\rm WP}\left(\varphi_{\rm WP}\right)$ in \eq{flowPolchinski2} by
a Legendre transform \cite{Morris:2005ck},
\begin{eqnarray}
v  &=& v_{\rm WP} +\frac{1}{2}\left(
\varphi_{\rm WP} -\varphi \right) ^{2} \,, \label{Legendre1} \\
\varphi  &=&\varphi_{\rm WP} -v_{\rm WP}^{\prime }\,, \label{Legendre2}
\end{eqnarray}
up to an irrelevant constant (the absolute value of the potentials at
vanishing field). An immediate consequence of \eq{Legendre2} is that the two
potentials take their absolute minimum at the same field value $\varphi_{{\rm
    WP},0}=\varphi_0$.  It follows that the barrier heights
\begin{equation}\label{barrier}
v(0)-v(\varphi_0)= v_{\rm WP}(0)-v_{\rm WP}(\varphi_{0})
\end{equation} 
are identical.  Further similarities and differences between \eq{flow} --
\eq{FP-Polchinski2} have been studied in \cite{Litim:2005us,Morris:2005ck}.

\section{Numerical methods}\label{Numerical}
In this section, we summarise the numerical techniques used to solve
\eq{flow} -- \eq{FP-Polchinski2} and similar to high accuracy at a
fixed point, and to deduce the leading and subleading scaling
exponents. These include tools for local polynomial expansions with
various degrees of sophistication (a) -- (c), tools for initial-value
problems (d), and tools for solving (two-point) boundary-value
problems (e).  Analytical methods for solving \eq{flow} can be found
in \cite{Tetradis:1995br,Litim:2002cf}, and are not further elaborated
here. \step

\subsection{Local behaviour}
For small fields, we use polynomial expansions of the effective potential
\cite{polynomial}. Their convergence has been addressed in
\cite{Morris:1994ki,Aoki:1996fn,Litim:2002cf}. We implement the expansion in
different ways (a) -- (c). At a fixed point, the potential is $Z_2$-symmetric
under $\varphi\to -\varphi$.  Therefore, we expand the effective potential in
$\rho$ rather than in $\varphi$. \step

(a) We Taylor-expand the potential $u(\rho)$ in field
monomials about vanishing field $\sim \rho^n$ up to the maximum order
$m$ in the truncation,
\begin{equation}\label{poly}
u(\rho)=\sum_{n=1}^{m}\frac{1}{n!}\lambda_n \rho^n\,.
\end{equation}
 Inserting \eq{poly} into \eq{flow} and solving for $\partial_t u'=0$
leads to coupled ordinary differential equations
$\partial_t\lambda_n=\beta_n(\{\lambda_i\})$ for the couplings
$\lambda_i$, which are solved numerically. We note that the $m$
differential equations $\partial_t\lambda_n$ depend, in general, on
$m+2$ couplings up to order $\lambda_{m+2}$. Therefore, we have to
provide boundary conditions for the couplings $\lambda_{m+1}$ and
$\lambda_{m+2}$, as their values are undetermined by the truncated
flow. For \eq{poly}, we employ $\lambda_{m+1}=0$ as a boundary
condition (in an expansion about vanishing field, the dependence on
$\lambda_{m+2}$ drops out), which shows good convergence with
increasing $m$; see Tab.~\ref{FPs} for numerical results for the
couplings at the fixed point of \eq{flow}.\step

(b) The convergence is further enhanced by expanding $u(\rho)$ about
non-vanishing field \cite{Aoki:1996fn,Litim:2002cf}. In this case, we
write
\begin{equation}\label{polyMin}
u(\rho)=\sum_{n=2}^{m}\frac{1}{n!}\lambda_n(\rho-\lambda_1)^n\,,
\end{equation} 
where the expansion point $\rho=\lambda_1$ is defined implicitly by
$u'(\lambda_1)=c$ with $c$ a free parameter. From \eq{flow} we
conclude that $c$ can take values between $-1$ and $\infty$. This is
confirmed by the explicit result, see Fig.~\ref{fUprime}. A natural
choice is $c=0$, in which case $\lambda_1$ denotes the unique
potential minimum, $u'(\lambda_1)=0$. The numerical convergence of the
expansion is best for small $c$. In a given truncation to order $m$,
the boundary condition $\lambda_{m+1}=0=\lambda_{m+2}$ is used. The
expansion displays very good convergence properties with increasing
$m$ \cite{Litim:2002cf}; see Tab.~\ref{FPs} for numerical results for
the couplings at the fixed point of \eq{flow}.\step

\begin{figure}
\begin{center}
\unitlength0.001\hsize
\begin{picture}(1000,910)
\put(230,630){\large $u(\rho)$}
\put(-15,880){\large $\infty$}
\put(20,10){\large -$\infty$}
\put(410,-50){\Large $\rho/\rho_0$}
\put(860,10){\large $\infty$}
\includegraphics[width=.9\hsize]{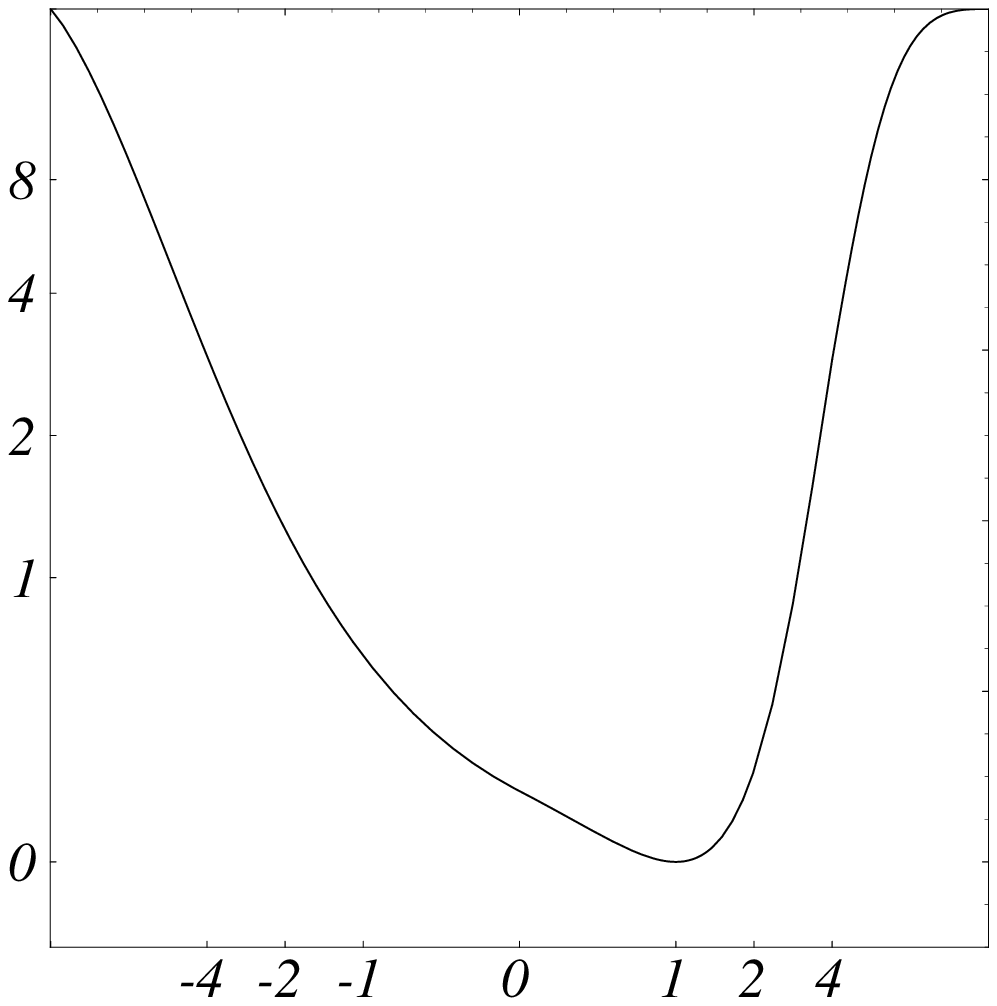}
\end{picture}
\vskip.5cm
\caption{\label{fUs} Fixed point potential $u(\rho)$ 
  from \eq{flow}, \eq{FP-opt} of the Ising universality class for all
  fields $\rho\in [-\infty,\infty]$. The absolute minimum is at
  $\rho_0=1.814\,898\,403\,687 \cdots$, where the potential is normalised to
  $u(\rho_0)=0$.  The axes are rescaled as
  $x\to\0{x}{2+|x|}$ with $x=\rho/\rho_0$,
  and $u\to\0{u}{2+u}$.}\vskip-.5cm
\label{A-la}
\end{center}
\end{figure}

(c) The simple boundary conditions $\lambda_{m+1}=0$ or
$\lambda_{m+1}=0=\lambda_{m+2}$ used in (a) or (b), respectively, fail to
provide convergent solutions for the Wilson-Polchinski flow
\eq{flowPolchinski}. The reason behind this is that the Wilson-Polchinski flow
at small fields is more strongly sensitive to the couplings at large fields
\cite{Litim:2005us}. Identifying a stable solution is then more demanding and
boils down to providing better-adapted boundary conditions for those higher
order couplings, which are undetermined in a given truncation. Equations which
fix the higher order couplings can be derived from the asymptotic behaviour of
the scaling solution.  An algebraic procedure which determines the highest
order couplings from the asymptotic behaviour of \eq{FP-opt}, $i.e.$ from
$u''\to 0$ for $\rho\to\infty$ [and similarly for \eq{FP-Polchinski}] has been
given in \cite{Boisseau:2006dq}. We have adopted this procedure for
\eq{FP-Polchinski} (for technical details, see \cite{Bervillier2007}). As a
result, we find that the polynomial expansion \eq{poly} for the
Wilson-Polchinski flow \eq{FP-Polchinski} with appropriate boundary condition
converges nicely; see Tab.~\ref{FPs-WP} for numerical results for the
couplings at the fixed point of \eq{flowPolchinski}. Although it is more
demanding to stabilise the small field expansion based on \eq{flowPolchinski}
as opposed to \eq{flow}, we note that the Wilson-Polchinski potential for
small fields differs only mildy from the optimised fixed point potential (see
Fig.~\ref{pComparisonSmallField} below). \step

    \begin{figure}
\begin{center}
\unitlength0.001\hsize
\begin{picture}(1000,910)
\put(230,630){\large $u'(\rho)$}
\put(10,310){\Large\rm -$\012$}
\put(10,180){\Large -$\045$}
\put(8,890){\large $\infty$}
\put(40,15){\large -$\infty$}
\put(440,-50){\Large $\rho/\rho_0$}
\put(890,10){\large $\infty$}
\includegraphics[width=.925\hsize]{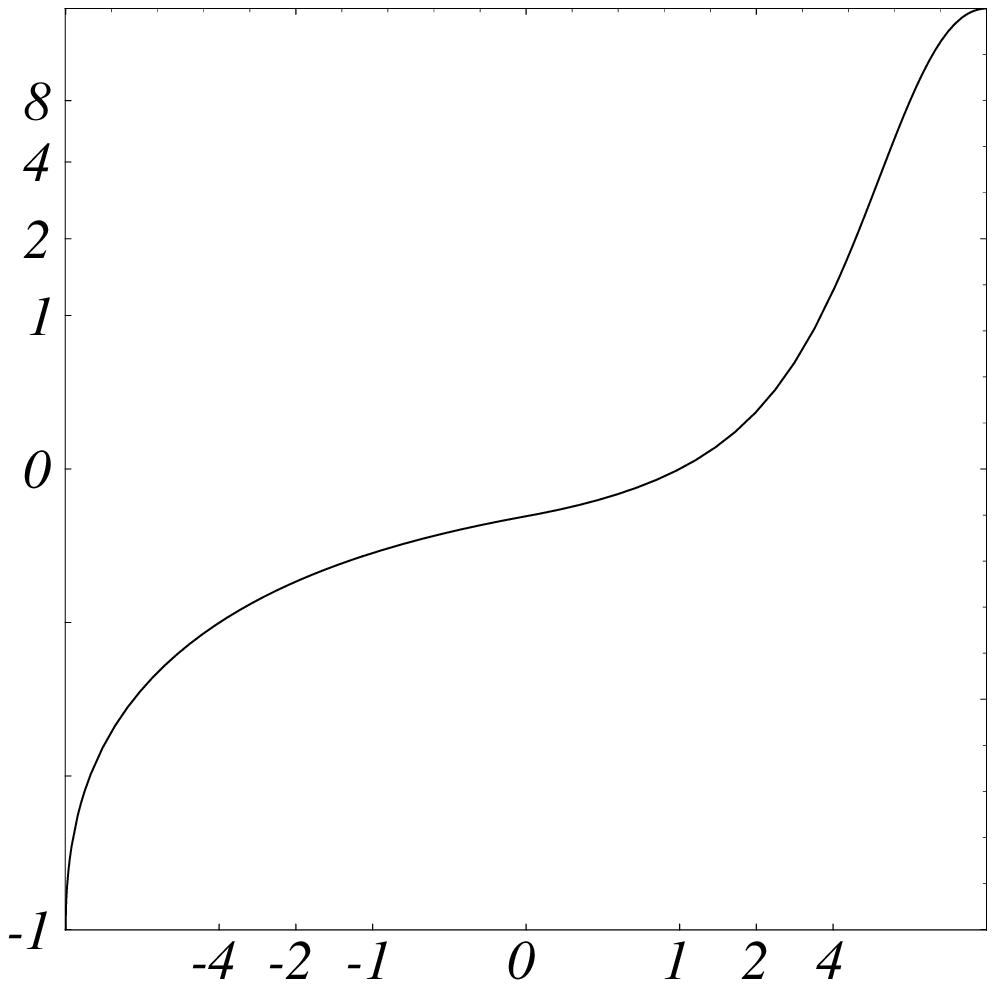}
\end{picture}
\vskip.5cm
\caption{\label{fUprime} First derivative of the fixed point potential
  $u'(\rho)$ from \eq{flow}, \eq{FP-opt} in the Ising universality class.
  Numerically, it reads $u'(0)=-0.186\,064\,249\,470 \cdots$ at vanishing
  argument. $u'$ is a monotonously increasing function of
the field. Same rescaling as in Fig.~\ref{fUs} (see text).}\vskip-.5cm
\label{A-la}
\end{center}
\end{figure}
Polynomial expansions like \eq{poly} and \eq{polyMin}
have, in general, a finite radius of convergence in field space,
restricted to field values $\rho_{\rm min}\le \rho\le\rho_{\rm
max}$. Here, we find that $\rho_{\rm min}<0$ and $\rho_{\rm
max}>\rho_0$. \step

In either of these cases (a) -- (c), the flows \eq{flow} and
\eq{flowPolchinski} transform into a set of $m$ coupled ordinary differential
equations $\partial_t\lambda_n=\beta_n(\{\lambda_i\})$ for the coefficient
functions $\lambda_n$. The fixed point equations $\beta_n(\{\lambda_i\})=0$
can be solved to very high accuracy with standard methods.  The scaling
exponent $\nu$ and subleading corrections-to-scaling exponents are deduced
from the eigenvalues of the stability matrix $M$ at criticality,
$M_{ij}=\partial_{\lambda_i} \beta_j|_{\partial_t u'=0}$. Asymmetric
corrections-to-scaling, $i.e.$ eigenperturbations not symmetric under
$\varphi\to-\varphi$, can be obtained as well \cite{Litim:2003kf}.

\begin{table}
\begin{center}
\begin{displaymath}
\begin{array}{c|ll}
{\rm coupling}& 
{}\quad\quad\quad
\rho=0
& 
\quad\ u'(\rho=\lambda_1)=0\\
\hline
\lambda_1
&               -0.186\,064\,249\,470
&\hspace{2.52mm} 1.814\,898\,403\,687 \\
\lambda_{2} 
& \hspace{2.52mm} 0.082\, 177\, 310\, 824
& \hspace{2.52mm} 0.126\, 164\, 421\, 218 \\
\lambda_{3} 
& \hspace{2.52mm} 0.018\, 980\, 341\, 099
& \hspace{2.52mm} 0.029\, 814\, 964\, 767 \\
\lambda_{4} 
& \hspace{2.52mm}  0.005\, 252\, 082\, 509
& \hspace{2.52mm}  0.006\, 262\, 816\, 384  \\
\lambda_{5} 
& \hspace{2.52mm} 0.001\, 103\, 954\, 106
&                -0.000\, 275\, 905\, 516
\end{array}
\end{displaymath}
\caption{The first five couplings at the fixed point of the optimised 
  flow \eq{flow} at vanishing field (first column), and at the potential
  minimum (second column).}\vskip-.5cm
\label{FPs}
\end{center}
\end{table}

\begin{table}
\begin{center}
\begin{displaymath}
\begin{array}{c|ll}
{\rm coupling}& 
{}\quad\quad\quad
\rho=0
& 
\quad\ u'(\rho=\lambda_1)=0\\
\hline
\lambda_1^{\rm WP}
&               -0.228\, 598\, 202 \, 437
&\hspace{2.52mm} 1.814\,898\,403\,687  \\
\lambda_{2}^{\rm WP} 
& \hspace{2.52mm} 0.187\, 236\, 893\, 730
& \hspace{2.52mm} 0.086\, 535\, 420\, 434 \\
\lambda_{3} ^{\rm WP}
&               -0.105\, 930\, 606\, 484
&                -0.028\, 253\, 169\, 622  \\
\lambda_{4} ^{\rm WP}
& \hspace{2.52mm}  0.101\, 611\, 201\, 027
& \hspace{2.52mm} 0.015\, 928\, 269\, 983 \\
\lambda_{5} ^{\rm WP}
&                -0.135\, 786\, 295\, 049 
&                -0.012\, 666\, 298\, 430            
\end{array}
\end{displaymath}
\caption{The first five 
  couplings at the fixed point of the Wilson-Polchinski flow
  \eq{flowPolchinski} at vanishing field (first column), and at the potential
  minimum (second column).}\vskip-.5cm
\label{FPs-WP}
\end{center}
\end{table}

\subsection{Global behaviour}

Next we discuss two methods (d) and (e) which integrate \eq{FP-opt} or
\eq{FP-Polchinski} directly, without using polynomial
approximations.\step

\begin{figure}
\begin{center}
\unitlength0.001\hsize
\begin{picture}(1000,910)
\put(230,630){\large $u'(\rho)+2 \rho\, u''(\rho)$}
\put(10,310){\Large -$\012$}
\put(10,180){\Large -$\045$}
\put(5,890){\large $\infty$}
\put(40,15){\large -$\infty$}
\put(440,-50){\Large $\rho/\rho_0$}
\put(890,10){\large $\infty$}
\includegraphics[width=.925\hsize]{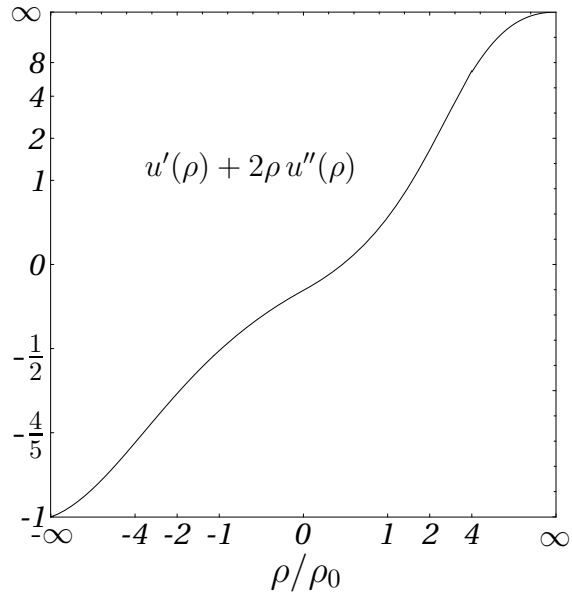}
\end{picture}
\vskip.65cm
\caption{\label{fUprime2} Mass function 
  $u'+2\rho\,u''$ of the fixed point potential \eq{FP-opt} in three
  dimensions, Ising universality class.  Same rescaling as in Fig.~\ref{fUs}
  (see text). For $\rho=\s012\varphi^2\ge 0$, the mass
  function is equivalent to $v''(\varphi)$ from \eq{FP-opt2}.}\vskip-.5cm
\label{A-la}
\end{center}
\end{figure}

(d) Initial value problem. The differential equations \eq{FP-opt} or
\eq{FP-Polchinski} are studied as initial value problems with boundary
conditions given at some starting point $\rho_1$. The necessary boundary
conditions can be obtained within the polynomial expansion (a) and (b) in
their domain of validity, see Tab.~\ref{FPs} for the couplings at vanishing
field and at the potential minimum. The boundary conditions are
\begin{eqnarray}\nonumber
u'(0)=\lambda_1\,,\ \ \, &&
u''(0)\equiv\lambda_2=-\s023\lambda_1(1+\lambda_1)^2\\
u'(0)=\lambda_1^{\rm WP}\,,&&
u''(0)\equiv\lambda_2^{\rm WP}=-\s023\lambda_1^{\rm WP}(1+\lambda_1^{\rm WP})
\nonumber
\end{eqnarray} 
at $\rho_1=0$ for the optimised RG flow \cite{Litim:2002cf} and the
Wilson-Polchinski flow, respectively. Identifying the well-defined fixed point
solution $u'(\rho)$ which extends over all fields $\rho$ requires a high
degree of fine-tuning in the boundary condition $\lambda_1$
\cite{Morris:1994ki}.  Integrating \eq{FP-opt} towards larger fields, starting
at some non-vanishing field $|\rho|\sim{\cal O}(1)$ with boundary conditions
from (a) or (b) is numerically more stable.  Following this strategy, we have
computed in Figs.~\ref{fUs}, \ref{fUprime} and \ref{fUprime2} the fixed point
potential $u$, its first derivative $u'$, and the field-dependent mass
term $u'+2\rho\,u''$, respectively, for all $\rho\in[-\infty,\infty]$.  We
note that $u$ displays only one global minimum. Both $u'$ and
$u'+2\rho\,u''$ are monotonously increasing functions of the field
variable. The numerical solution fits the expected analytical behaviour for
asymptotically large fields. \step

(e) Finally, we discuss solvers for two-point boundary value problems. The
shooting method \cite{Recipes} requires boundary conditions at two points
$\rho_1$ and $\rho_2$ in field space.  Initial conditions at $\rho_1$ are
varied until the boundary condition at $\rho_2$ is matched. The procedure is
iterated until the desired accuracy in the solution is achieved
\cite{Recipes}. If one is not constrained by the $Z_2$ symmetry
$\varphi\to-\varphi$ and because of the potentially singular behaviour of the
right-hand sides of \eq{FP-opt} and \eq{FP-Polchinski} for $\rho\to 0$, it is
preferable to implement \eq{FP-opt2} and \eq{FP-Polchinski2}. Then the
differential equations at the origin $\varphi=0$ are better under control and
one may shoot from the origin $\varphi_1=0$ (with the required symmetry
conditions imposed) to large fields $\varphi_2\approx\varphi_{\rm bound}$,
where the asymptotic large-field behaviour is imposed. Shooting in the reverse
direction may provide a better accuracy. The large-field behaviour of
\eq{FP-opt}, \eq{FP-Polchinski} and \eq{FP-opt2}, \eq{FP-Polchinski2} has
previously been determined in the literature, $e.g.$~\cite{Litim:2005us}. The
RG eigenvalues are deduced from \eq{flow} or \eq{flowPolchinski} in the
vicinity of the fixed point solution. The implementation of \eq{flow2},
\eq{FP-opt2} and \eq{flowPolchinski2}, \eq{FP-Polchinski2} in terms of
$\varphi$ allows a direct computation of asymmetric correction-to-scaling
exponents.

    \begin{figure}
\begin{center}
\unitlength0.001\hsize
\begin{picture}(1000,0)
\includegraphics[angle=270,width=\hsize]{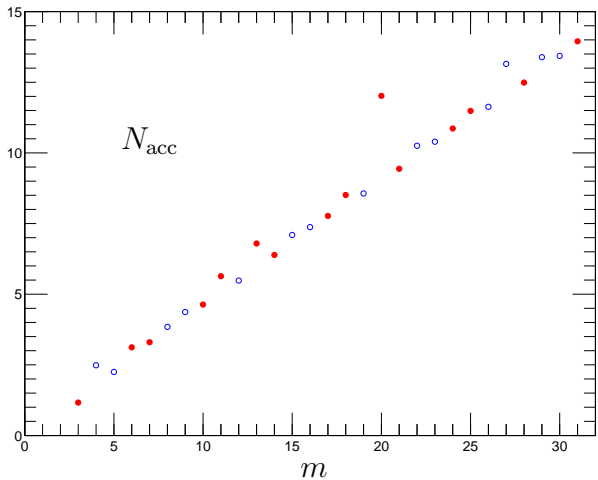}
\put(-800,-250){\large $N_{\rm acc}$}
\put(-500,-800){\large $m$}
\end{picture}
\vskip6.5cm
\caption{\label{fAccuracy} Accuracy of the fixed point, as defined in
 \eq{acc}, and its dependence on the order of the expansion $m$; from
 \eq{flow} using method (b) with $c=0$.}\vskip-.5cm
\label{A-la}
\end{center}
\end{figure}

\section{Error control}\label{Error}
Within the numerical approaches (a) -- (e), there are several sources
for numerical errors, the control of which is discussed in this
section.\step

(i) Within the polynomial expansions (a) -- (c), the fixed point is
determined by seeking simultaneous zeros of all
$\beta$-functions. Solutions for $\partial_t\lambda_n=0$ are found to
very high precision. To ensure that the fixed point is numerically a
good approximation to the full fixed point solution, we have computed
$|\partial_t u'(\rho)|$ for all $\rho$ within the domain of validity
 of the polynomial approximation. This serves as a measure for the
quality of a polynomial expansion to order $m$. We define the accuracy
$N_{\rm acc}$ of the fixed point solution to order $m$ as
\begin{equation}\label{acc}
10^{-N_{\rm acc}}=
\max_{\rho\in [0,\rho_{\rm max}]} |\partial_t u'(\rho)| 
\,.
\end{equation}
Here, $[0,\rho_{\rm max}]$ denotes a compact neighbourhood of the expansion
point $\lambda_1$, which needs not to coincide with the potential minimum
$\rho_0$. We take $\rho_{\max}>\rho_0$ from $u'(\rho_{\rm max})+2\rho_{\rm
  max}\,u''(\rho_{\rm max})=1$. This notion of accuracy is a good measure for
how well a local Taylor expansion to order $m$ approximates the full fixed
point potential on $[0,\rho_{\rm max}]$.  Using (b) with $c=0$, we find that
\eq{acc} achieves its extremum typically at $\rho=0$. In Fig.~\ref{fAccuracy},
we display $N_{\rm acc}$ as a function of $m$.  Full red (open blue) dots
indicate that $\max \partial_t u'$ is positive (negative). The slope is
approximately $1/2$, indicating that an increase in the truncation by $\Delta
m\approx 2$ increases the accuracy in the fixed point solution roughly by a
decimal place. This serves as an indicator for the reliability in the scaling
exponents [see (ii)]. \step

    \begin{figure}
\begin{center}
\unitlength0.001\hsize
\begin{picture}(1000,0)
\includegraphics[angle=-90,width=\hsize]{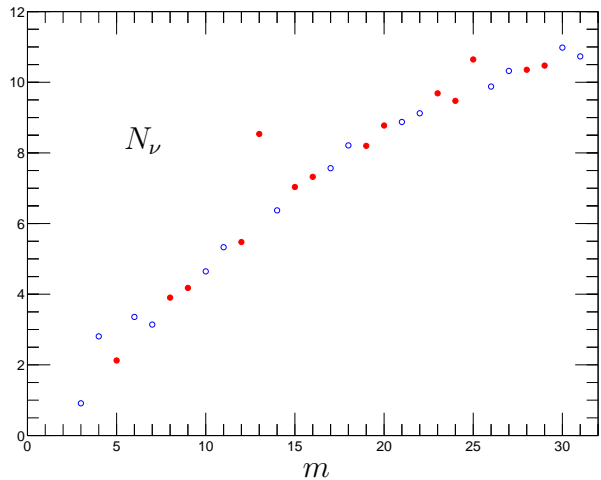}
\put(-800,-250){\large $N_{\nu}$}
\put(-500,-800){\large $m$}
\end{picture}
\vskip6.5cm
\caption{\label{fAccuracy-nu} Accuracy of the scaling exponent $\nu$,
  as defined by \eq{acc-nu}, and its dependence on the order of the
  expansion $m$; from \eq{flow} using method (b) with $c=0$. }\vskip-.5cm
\label{A-la}
\end{center}
\end{figure}

(ii) Within the polynomial expansions (a) -- (c), we study the
numerical convergence of both the fixed point values and the scaling
exponents with increasing degree of truncation $m$
\cite{Litim:2002cf}. In analogy to \eq{acc}, we define the number of
significant figures $N_X$ in a fixed point coupling or a critical
index $X_m$ at order $m$ in the expansion as
\begin{equation}\label{acc-nu}
10^{-N_{X}}= \left|1-\frac{X_m}{X}\right|\,,
\end{equation}
where $X$ denotes the full (asymptotic) result. For the leading scaling
exponent $X=\nu$, we display $N_\nu$ as a function of $m$ in
Fig.~\ref{fAccuracy-nu}, based on the expansion \eq{polyMin} with $c=0$. An
open blue (full red) dot indicates that $\nu(m)$ is smaller (larger) than the
asymptotic value. The expansion converges roughly in the pattern $++--$.  Note
that the slope, in comparison with Fig.~\ref{fAccuracy}, slightly decreases
towards larger $m$. This part of the analysis is conveniently performed with
{\sc Mathematica}. The accuracy of the matrix inversion (leading to the
scaling exponents) has been checked independently with standard routines from
{\sc Matlab}.\step

\begin{figure}
\begin{center}
  \unitlength0.001\hsize
\begin{picture}(1000,950)
\includegraphics[width=\hsize]{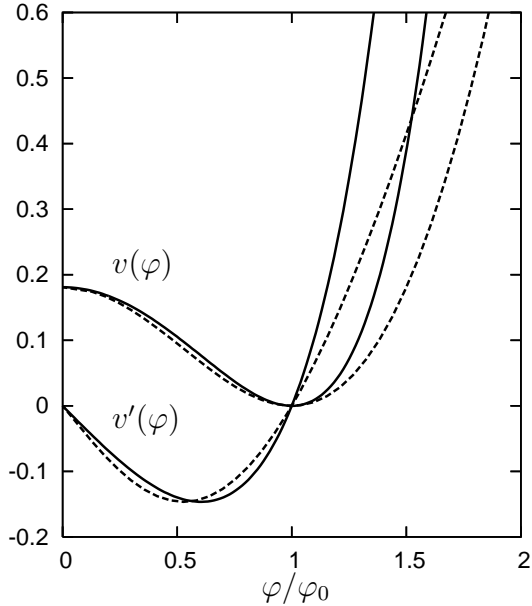}
\put(-500,-20){\large $\varphi/\varphi_0$}
\put(-750,520){\large $v(\varphi)$}
\put(-750,260){\large $v'(\varphi)$}
\end{picture}
\vskip.2cm
\caption{\label{OWP} Comparison of the fixed point potentials $v(\varphi)$ 
  and their first derivatives $v'(\varphi)$ from the optimised RG (full lines)
  and the Wilson-Polchinski RG (dashed lines). Both potentials are normalised
  to $v(\varphi_0)=0$, where they take their absolute minimum. In either case,
  $\varphi_0=\sqrt{2\rho_0}=1.905\,202\,563\,344\cdots$.}\vskip-.5cm
\label{pComparisonSmallField}
\end{center}
\end{figure}

(iii) Within the polynomial expansions (a) and (b), we study the
numerical stability of the result by varying the expansion point. The
radius of convergence of polynomial expansions depends on the
latter. This check serves as an indicator for a possible break-down of
the expansion at the highest orders. We have confirmed that only the
rate of convergence depends on the expansion point, but not the
asymptotic result.\step

(iv) The numerical approach (d) is checked in several ways. First, starting at
intermediate fields, the numerical accuracy in the integration can be made
large. Second, the domain of validity for (d) and (a) -- (c) overlap. This
allows for a quantitative cross-check. Third, for large fields
$|\rho|\to\infty$, the results from (d) match the expected asymptotic
behaviour. Fourth, we have cross-checked the result with a two-point boundary
value routine.\step

(v) The shooting method (e) is controlled in several ways. First, the
numerical precision for the solution on $[0,\rho_{\rm bound}]$ is only limited
by the machine precision. We use standard routines under {\tt Fortran f77}
with double precision. The boundary condition at large fields has not to be
known to very high accuracy to achieve a reliable result.  Stability in the
result is confirmed by varying $\rho_{\rm bound}$ as well as the boundary
condition at $\rho_{\rm bound}$. These procedures are applied for both
\eq{FP-opt} and \eq{FP-Polchinski}. The shooting method is checked
independently using a different solver under {\sc Maple}, based on a different
boundary value integration algorithm.

\begin{figure}
\begin{center}
\unitlength0.001\hsize
\begin{picture}(1000,910)
\put(130,330){\large $v(\varphi)$}
\put(400,800){\large $\sim \varphi^6$}
\put(700,800){\large $\sim \varphi^2$}
\put(-15,880){\large $\infty$}
\put(410,-50){\large $\varphi/\varphi_0$}
\put(860,10){\large $\infty$}
\includegraphics[width=.9\hsize]{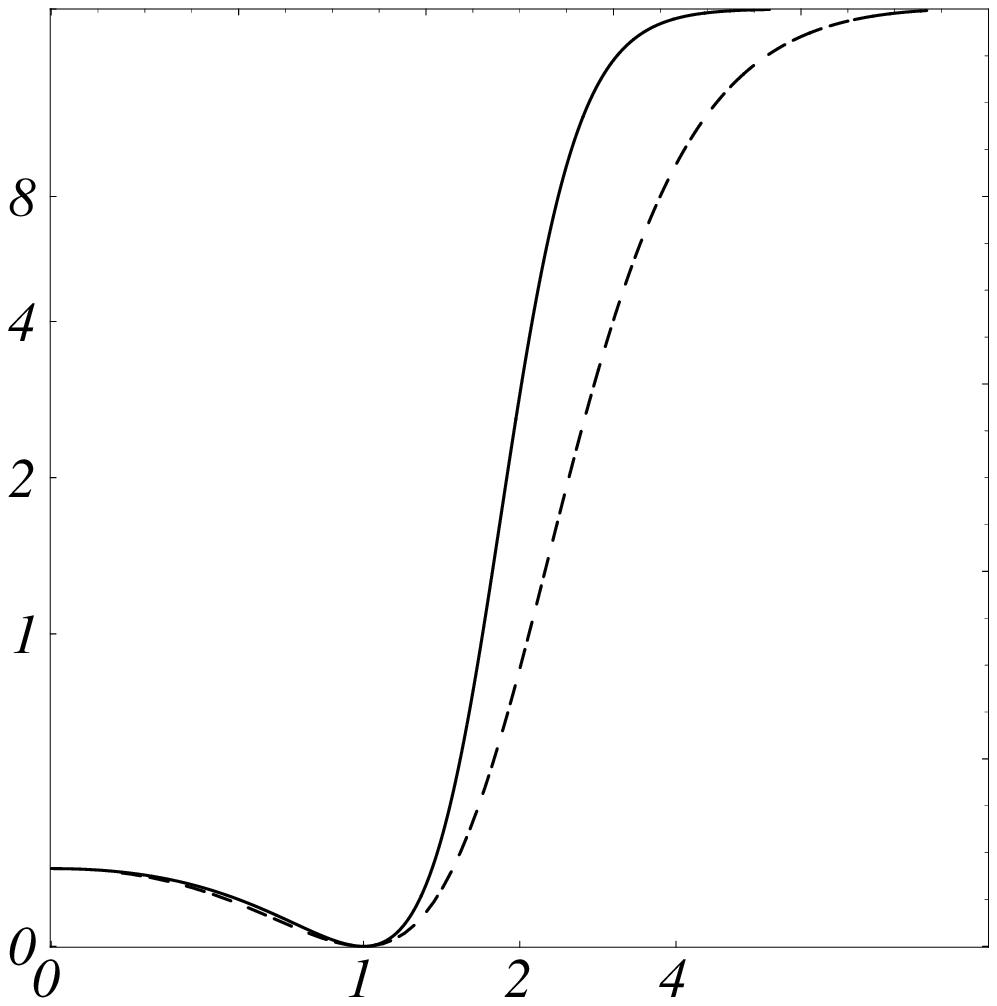}
\end{picture}
\vskip.5cm
\caption{\label{pComparisonLargeField} Comparison of 
  the fixed point potentials $v(\varphi)$ from the optimised RG (full line)
  and the Wilson-Polchinski RG (dashed line) for all fields $\varphi\in
  [0,\infty]$. As in Fig.~\ref{pComparisonSmallField}, the potentials are
  normalised to $v(\varphi_0)=0$. For display purposes, the axes are rescaled
  as $x\to\0{x}{2+|x|}$ with $x=\varphi/\varphi_0$, and
  $v\to\0{v}{2+v}$.}\vskip-.5cm
\end{center}
\end{figure}

\section{Results}\label{Results}
Our results for the scaling exponents from \eq{flow} and \eq{flowPolchinski2}
are given in Tab.~\ref{tSymmetric} and \ref{tAsymmetric}. All digits are
significant, except for possible rounding effects. Earlier findings
with a lower level of accuracy
\cite{Comellas:1998ep,Litim:2002cf,Bervillier:2004mf} are fully consistent
with Tab.~\ref{tSymmetric}. Specifically, in Tab.~\ref{tSymmetric}, the
exponents from the optimised RG flows have been obtained using (a), (b), and
(e), whereas those for the Wilson-Polchinski flow follow from (c) and (e). The
method (b) has been used up to the order $m=32$, leading to roughly 11
significant digits in the result for $\nu$, and roughly one significant digit
less with increasing order for the subleading exponents,
see~\cite{Litim:2002cf,Litim:2003kf}. Method (e) has been pushed to an
accuracy of 14 digits for all scaling exponents. The asymmetric
corrections-to-scaling exponents in Tab.~\ref{tAsymmetric} have been computed
using (e), again to an accuracy of 14 digits. Again, all earlier results to
lower order in the accuracy \cite{Litim:2003kf,Bervillier:2004mf} are
consistent with Tab.~\ref{tAsymmetric}.\step

In Figs.~\ref{fUs}~--~\ref{fUprime2}, we have plotted our results for the
fixed point potential $u(\rho)$ from \eq{FP-opt} and its derivatives for all
$\rho\in[-\infty,\infty]$. It is noteworthy that the solution extends to all
negative $\rho$, which in terms of the field $\varphi$ corresponds to the
analytical continuation to purely imaginary field values. For
$\rho\to-\infty$, the solution approaches the convexity bound where $u'+2\rho
u''\to-1$. In the flow equation \eq{flow}, this limit is potentially singular
as it corresponds to a pole.  However, the explicit solution for $\partial_t
u'=0$ shows that the pole is supressed, leading to a well-behaved solution
even in this limit. In the physical domain where $\rho\ge 0$, the fixed point
solution stays far away from the pole, since $\min_{\rho\ge 0}(u'+2\rho
u'')=u'(0)>-1$. General (non-fixed point) solutions to \eq{flow} approach the
pole only in a phase with spontaneous symmetry breaking where $V_k''(\phi)$ is
negative. Then the pole ensures convexity of the physical potential
$V_{k=0}(\phi)$ in the infrared limit where $V_{k=0}''(\phi)\ge0$
\cite{Litim:2006nn}.\step

In Figs.~\ref{pComparisonSmallField} and~\ref{pComparisonLargeField}, we
compare the fixed point potentials $v(\varphi)$ from the optimised RG and the
Wilson-Polchinski RG for small and large fields. We have checked that
$u(\rho)$ in Fig.~\ref{fUs} matches $v(\varphi)$ in
Fig.~\ref{pComparisonLargeField}, as it should. Both potentials have their
absolute minimum at the same field value $\rho_0^{\rm WP}=\rho_0
=1.814\,898\,403\,687 $; see Tab.~\ref{FPs} and~\ref{FPs-WP}. It is also
confirmed that the potentials display the same barrier height
$v(0)-v(\varphi_0)$ as expected from \eq{barrier}. For large fields, the
potentials scale differently with the fields \cite{Litim:2005us}. \step

\begin{table}
\begin{center}
\begin{displaymath}
\begin{array}{c|ll}
{\rm exponent}
& {}\quad{\rm optimised\ RG}
& {\rm Wilson}{\rm \small -}{\rm Polchinski\ RG} \\
\hline
\nu 
& {\ \,   0.649\,561\,773\,880\,65\ } 
& {\ \ \, 0.649\,561\,773\,880\,65} \\
\omega  
& {\ \,   0.655\,745\,939\,193\,3\ } 
& {\ \ \, 0.655\,745\,939\,193\,35} \\
\omega _{2} 
& {\ \,   3.180\,006\,512\,059\,2\ } 
& {\ \ \, 3.180\,006\,512\,059\,2} \\
\omega _{3}
& {\ \,   5.912\,230\,612\,747\,7\ } 
& {\ \ \, 5.912\,230\,612\,747\,7} \\
\omega _{4}
& {\ \,   8.796\,092\,825\,414\ } 
& {\ \ \, 8.796\,092\,825\,414} \\
\omega _{5}
& {  11.798\,087\,658\,337} 
& {\ 11.798\,087\,658\,336\,9} \\
\omega _{6}
& {  14.896\,053\,175\,688} 
& {\ 14.896\,053\,175\,688}
\end{array}
\end{displaymath}
\caption{The leading scaling exponent $\nu$ and the first six
    subleading eigenvalues for the Ising universality class
    in three dimensions.}\vskip-.5cm
\label{tSymmetric}
\end{center}
\end{table}

We have checked numerically that the fixed point solutions to \eq{FP-opt} and
\eq{FP-Polchinski} are related by a Legendre transform, see \eq{Legendre1},
\eq{Legendre2}. Consequently, at vanishing field, the couplings from
Tab.~\ref{FPs} are related to the Wilson-Polchinski couplings in
Tab.~\ref{FPs-WP} by
\begin{eqnarray}
\lambda_1=\frac{\lambda_1^{\rm WP}}{1-\lambda_1^{\rm WP}}\,,&&
\lambda_2=\frac{\lambda_2^{\rm WP}}{(1-\lambda_1^{\rm WP})^4}\,, \\
\lambda^{\rm WP}_1=\frac{\lambda_1}{1+\lambda_1}\,,&&
\lambda^{\rm WP}_2=\frac{\lambda_2}{(1+\lambda_1)^4}\,.
\end{eqnarray}
Similar expressions are found for the higher order couplings. The numerical
results in Tab.~\ref{FPs} and~\ref{FPs-WP} confirm these relations within the
present accuracy. Our value $\lambda_1^{\rm WP}=-0.228\, 598\, 202 \, 437\,
02$ for the dimensionless mass term squared at vanishing field deviates at the
order $10^{-5}$ from the corresponding value $-0.228\, 601\,293\,102$ given in
\cite{Ball:1994ji}. \step

In summary, we have confirmed the equivalence of an optimised RG and
the Wilson-Polchinski RG to the order $10^{-14}$ in the universal 
indices, and in the Legendre-transformed fixed point solutions.

\begin{table}[t]
\begin{center}
\begin{displaymath}
\begin{array}{c|ll}
{\rm exponent}
& {}\quad{\rm optimised\ RG}
& {\rm Wilson}{\rm \small -}{\rm Polchinski} \\
\hline
\bar{\omega} 
& {\ \,  1.886\,703\,838\,091\,4}
& {\ \ \,1.886\,703\,838\,091\,4} \\
\bar{\omega}_{2} 
& {\ \,  4.524\,390\,733\,670\,8} 
& {\ \ \,4.524\,390\,733\,670\,8} \\
\bar{\omega}_{3} 
& {\ \,  7.337\,650\,643\,354} 
& {\ \ \,7.337\,650\,643\,354\,4} \\
\bar{\omega}_{4} 
&   10.283\,900\,724\,026 
& \ 10.283\,900\,724\,026%
\end{array}
\end{displaymath}
\caption{The first four asymmetric correction-to-scaling exponents
    for the Ising universality class in three dimensions.}\vskip-.5cm
\label{tAsymmetric}
\end{center}
\end{table}

\section{Comparison}\label{Comparison}

Next we compare our results with those based on Dyson's hierarchical
model \cite{Dyson:1968up}. To leading order in the derivative
expansion, it has been proven by Felder \cite{Felder:1987} that the
hierarchical RG is equivalent to the Wilson-Polchinski RG, and hence to the
optimised RG. Then scaling exponents should come out identical.  With
high-accuracy numerical data at hand, we can test this assertion
quantitatively.\step

In hierarchical models, the block-spin transformations are 
characterised by a decimation parameter $\ell\ge 1$, where $\ell\to 1$ refers
to Felder's limit of continuous transformations.
Numerical studies with
$\ell=2^{1/3}\approx1.26$ have been performed in
\cite{Wilson:1972,Pinn:1994st,Godina:1997dk,Godina:1998nh,Godina:1998uz,Gottker-Schnetmann:1999eg,Godina:2005hv}.
In Tab.~\ref{tComparisonNu}, we compare our results with previous computations
of $\nu$ from the Wilson-Polchinski RG \cite{Ball:1994ji,Comellas:1997tf}, the
optimised RG \cite{Litim:2001fd,Litim:2002cf}, and the hierarchical RG
\cite{Wilson:1972,Pinn:1994st,Godina:1997dk,Godina:1998nh,Godina:1998uz,Gottker-Schnetmann:1999eg,Godina:2005hv}.
The most advanced hierarchical RG (HRG) and functional RG (FRG) studies agree
amongst themselves at least to the order $10^{-6}$ and $10^{-12}$,
respectively. We emphasize that the functional RG results clearly deviate from
the hierarchical RG results, although the difference $(\nu_{\rm FRG}-\nu_{\rm
  HRG})/\nu_{\rm FRG}\approx -1.3\times 10^{-5}$ is quantitatively small.
\step

\begin{table}
\begin{center}
\begin{displaymath}
\begin{array}{lll}
{\rm Wilson}-{\rm Polchinski} 
& 
\  {\rm optimised\ RG}\ 
& 
{\rm hierarchical\ RG}\ \\
\hline
  0.649^c 
& 
  0.64956^g 
& 0.6496^a  
\\ 
0.6496^d 
& 
0.649562^h 
&
0.64957^b 
\\
  0.649561773881{}
& 0.649561773881{}
& 0.649570^{e,f,i} 
\end{array}
\end{displaymath}
\caption{Comparison of the scaling exponent $\nu$ for the Ising
  universality class in three dimensions (see text). Data from this work, and
  from $a)$ \cite{Wilson:1972}, $b)$~\cite{Pinn:1994st}, $c)$
  \cite{Ball:1994ji}, $d)$ \cite{Comellas:1997tf}, $e)$ \cite{Godina:1998nh},
  $f)$ \cite{Gottker-Schnetmann:1999eg}, $g)$ \cite{Litim:2001fd}, $h)$
  \cite{Litim:2002cf}, $i)$ \cite{Godina:2005hv}.}\vskip.5cm
\label{tComparisonNu}
\end{center}
\end{table}

In Tab.~\ref{tComparison} we compare the best results from the functional RG
($i.e.$~both the Wilson-Polchinski RG and the optimised RG) with those
from the hierarchical RG for various other indices, some of which have been
obtained with up to 12 significant figures~\cite{Godina:1998nh} (see also
\cite{Godina:2005hv}). The variations $(X_{\rm FRG}-X_{\rm HRG})/X_{\rm FRG}$
in the exponents $X=\nu,\omega,\gamma,\Delta, \alpha$ read $(-1.3, 1.5, -1.3,
0.15, 5.0) \times 10^{-5}$, respectively. Hence, we confirm that the findings
disagree, beginning at the order $10^{-5}$. Given Felder's proof of
equivalence for the limit $\ell\to 1$ \cite{Felder:1987}, we conclude that the
hierarchical model diplays a dependence on the decimation parameter $\ell$,
with a tiny slope of the order of $10^{-6}$ for variations in the exponents.
\step

\section{Discussion and conclusions}\label{Conclusions}

Equivalences between non-linear differential or difference equations, and more
generally functional relationships between different implementations of the
renormalisation group, often allow for a deeper understanding of both the
underlying physics and the adequacy and efficiency of the methods at hand.  It
has been proven previously that three different implementations of Wilson's
renormalisation group -- the Wilson-Polchinski RG, an optimised version of the
effective average action RG, and Dyson's hierarchical RG in the limit of
continuous block-spin transformations -- are equivalent in the local potential
approximation. This implies that the corresponding non-linear RG equations
carry identical universal content, $e.g.$~identical scaling exponents.\step

Quantitatively, these equivalences have been confirmed up to the order
$10^{-4}$ in the literature, but a discrepancy at the order $10^{-5}$ has
recently been pointed out. This last observation relies on studies within
an optimised RG and Dyson's hierarchical RG with decimation parameter
$\ell=2^{1/3}$, where results with a sufficiently high accuracy are available,
while previous results from the Wilson-Polchinski RG were in agreement with
either of them. Here, we have closed this gap in the literature by computing
scaling exponents for the Ising universality class from the Wilson-Polchinski
RG and the optimised RG with an unprecedented accuracy of the order
$10^{-14}$. We confirm their equivalence, first conjectured in
\cite{Litim:2001fd}, based on the leading, subleading and asymmetric
correction-to-scaling exponents. Equally important, our central numerical
results are obtained in several ways, and furthermore independently from both
the Wilson-Polchinski and the optimised RG flow. We conclude that their
equivalence is rock solid for all technical purposes, and in full agreement
with the explicit proof given by Morris \cite{Morris:2005ck}.\step

\begin{table}[t]
\begin{center}
\begin{displaymath}
\begin{array}{c|ll}
{\rm exponent} 
& {}\quad\quad{\rm functional\ RG}
& {}\quad{\rm hierarchical\ RG}\\
\hline
\nu 
& {\ \,   0.649\,561\,773\,880\,65\ } 
& {\ \ \, 0.649\,570\, {}^{e,f,i}} \\
\omega  
& {\ \,   0.655\,745\,939\,193\,3\ } 
& {\ \ \, 0.655\,736\, {}^{i}} \\
\gamma 
& {\ \,   1.299\, 123\, 547\, 761\, 3\ } 
& {\ \ \, 1.299\, 140\, 730\, 159{}^{e}} \\
\Delta
& {\ \,   0.425\, 947\, 495\, 477\, 4 } 
& {\ \ \, 0.425\, 946\, 859\, 881{}^{e}} \\
\alpha
& {\ \,   0.051\, 314\, 678\, 358\,05\ } 
& {\ \ \, 0.051\, 289\, {}^{i}} \\
\end{array}
\end{displaymath}
\caption{Comparison of universal indices $\nu$, $\omega$, 
  $\gamma=2 \,\nu$, $\Delta = \omega\,\nu$ and $\alpha=2-3\nu$ (for $\eta=0$)
  from the functional RG (this work) and the hierarchical RG 
  (same referencing as in Tab.~\ref{tComparisonNu}).}\vskip-.5cm
\label{tComparison}
\end{center}
\end{table}
In contrast to this, we now have clear indications for a non-equivalence of
our results with those from Dyson's hierarchical model for discrete block-spin
transformations. In the $3d$ Ising universality class, all scaling exponents
from the functional RG, $i.e.$ from both the Wilson-Polchinski and the
optimised RG, differ systematically at the order $10^{-5}$ from high accuracy
studies based on Dyson's hierarchical RG.  Given the high degree of accuracy
in the critical indices from the present and earlier studies, and the fact
that the scaling exponents in all three approaches have been obtained from
several independent numerical implementations and collaborations, it is
unlikely that the discrepancy originates from numerical insufficiencies.
Furthermore, these differences are absent in the limit of continuous
block-spin transformations \cite{Felder:1987}. We conclude that the tacit
assumption of $\ell$-independence in the scaling exponents of the hierarchical
model \cite{Godina:2005hv} cannot be maintained.  Rather, the hierarchical
model carries an inherent $\ell$-dependence, similar to a dependence
previously observed in \cite{Meurice:1996bh} (see also \cite{Meurice:2007zg}).
In functional RG approaches, scheme-dependences related to the
underlying Wilsonian momentum cutoff can arise in truncations
\cite{Bagnuls:2000ae}.
Here, powerful control and optimisation mechanisms are available which
decrease truncational artefacts and increase convergence towards the physical
theory \cite{Liao:1999sh,Litim:2001up,Pawlowski:2005xe}.  It will be
interesting to see if the $\ell$-dependence of hierarchical models or its
extensions can be understood along similar lines.\step

For our numerical work, we have developed methods to solve non-linear
eigenvalue problems with high precision. The combined use of local polynomial
expansions techniques with global integration methods allows for an efficient
determination of the full fixed point solutions with reasonable numerical
efforts.  We have also put some emphasis on a reliable error control, in view
of the high accuracy aimed for in the universal eigenvalues.  We expect that
this combination of techniques will be equally useful for other non-linear
problems in mathematical physics, including $e.g.$~the derivative expansion to
higher order.\step

{\it Acknowledgements.} We thank Humboldt University for computer time.  AJ is
supported by PPARC grants PPA/G/S/2002/00467 and PPA/G/O/2002/0046.  DFL is
supported by an EPSRC Advanced Fellowship.

\end{document}